\documentstyle[twoside,psfig]{article}

\catcode`\@=11
\long\def\@makefntext#1{
\protect\noindent \hbox to 3.2pt {\hskip-.9pt  
$^{{\eightrm\@thefnmark}}$\hfil}#1\hfill}		

\def\thefootnote{\fnsymbol{footnote}}
\def\@makefnmark{\hbox to 0pt{$^{\@thefnmark}$\hss}}	
	
\def\ps@myheadings{\let\@mkboth\@gobbletwo
\def\@oddhead{\hbox{}
\rightmark\hfil\eightrm\thepage}   
\def\@oddfoot{}\def\@evenhead{\eightrm\thepage\hfil
\leftmark\hbox{}}\def\@evenfoot{}
\def\sectionmark##1{}\def\subsectionmark##1{}}

\oddsidemargin=\evensidemargin
\renewcommand{\thefootnote}{\fnsymbol{footnote}}

\newcounter{sectionc}\newcounter{subsectionc}\newcounter{subsubsectionc}
\renewcommand{\section}[1] {\vspace{12pt}\addtocounter{sectionc}{1} 
\setcounter{subsectionc}{0}\setcounter{subsubsectionc}{0}\noindent 
	{\tenbf\thesectionc. #1}\par\vspace{5pt}}
\renewcommand{\subsection}[1] {\vspace{12pt}\addtocounter{subsectionc}{1} 
	\setcounter{subsubsectionc}{0}\noindent 
	{\bf\thesectionc.\thesubsectionc. {\kern1pt \bfit #1}}\par\vspace{5pt}}
\renewcommand{\subsubsection}[1] {\vspace{12pt}\addtocounter{subsubsectionc}{1}
	\noindent{\tenrm\thesectionc.\thesubsectionc.\thesubsubsectionc.
	{\kern1pt \tenit #1}}\par\vspace{5pt}}
\newcommand{\nonumsection}[1] {\vspace{12pt}\noindent{\tenbf #1}
	\par\vspace{5pt}}

\newcounter{appendixc}
\newcounter{subappendixc}[appendixc]
\newcounter{subsubappendixc}[subappendixc]
\renewcommand{\thesubappendixc}{\Alph{appendixc}.\arabic{subappendixc}}
\renewcommand{\thesubsubappendixc}
	{\Alph{appendixc}.\arabic{subappendixc}.\arabic{subsubappendixc}}

\renewcommand{\appendix}[1] {\vspace{12pt}
        \refstepcounter{appendixc}
        \setcounter{figure}{0}
        \setcounter{table}{0}
        \setcounter{lemma}{0}
        \setcounter{theorem}{0}
        \setcounter{corollary}{0}
        \setcounter{definition}{0}
        \setcounter{equation}{0}
        \renewcommand{\thefigure}{\Alph{appendixc}.\arabic{figure}}
        \renewcommand{\thetable}{\Alph{appendixc}.\arabic{table}}
        \renewcommand{\theappendixc}{\Alph{appendixc}}
        \renewcommand{\thelemma}{\Alph{appendixc}.\arabic{lemma}}
        \renewcommand{\thetheorem}{\Alph{appendixc}.\arabic{theorem}}
        \renewcommand{\thedefinition}{\Alph{appendixc}.\arabic{definition}}
        \renewcommand{\thecorollary}{\Alph{appendixc}.\arabic{corollary}}
        \renewcommand{\theequation}{\Alph{appendixc}.\arabic{equation}}
        \noindent{\tenbf Appendix \theappendixc #1}\par\vspace{5pt}}
\newcommand{\subappendix}[1] {\vspace{12pt}
        \refstepcounter{subappendixc}
        \noindent{\bf Appendix \thesubappendixc. {\kern1pt \bfit #1}}
	\par\vspace{5pt}}
\newcommand{\subsubappendix}[1] {\vspace{12pt}
        \refstepcounter{subsubappendixc}
        \noindent{\rm Appendix \thesubsubappendixc. {\kern1pt \tenit #1}}
	\par\vspace{5pt}}

\newcommand{\textlineskip}{\baselineskip=13pt}
\newcommand{\smalllineskip}{\baselineskip=10pt}

\def\eightcirc{
\begin{picture}(0,0)
\put(4.4,1.8){\circle{6.5}}
\end{picture}}
\def\eightcopyright{\eightcirc\kern2.7pt\hbox{\eightrm c}} 

\newcommand{\copyrightheading}[1]
	{\vspace*{-2.5cm}\smalllineskip{\flushleft
	{\footnotesize International Journal of Modern Physics A #1}\\
	{\footnotesize $\eightcopyright$\, World Scientific Publishing
	 Company}\\
	 }}

\newcommand{\publisher}[2]{{\begin{center}\footnotesize\smalllineskip 
	Received #1\\
	Revised #2
	\end{center}
	}}

\def\abstracts#1#2#3{{
	\centering{\begin{minipage}{4.5in}\footnotesize\baselineskip=10pt
	\parindent=0pt #1\par 
	\parindent=15pt #2\par
	\parindent=15pt #3
	\end{minipage}}\par}} 

\newcommand{\bibit}{\nineit}

\renewenvironment{thebibliography}[1]
	{\frenchspacing
	 \ninerm\baselineskip=11pt
	 \begin{list}{\arabic{enumi}.}
	{\usecounter{enumi}\setlength{\parsep}{0pt}
	 \setlength{\leftmargin 12.7pt}{\rightmargin 0pt} 
	 \setlength{\itemsep}{0pt} \settowidth
	{\labelwidth}{#1.}\sloppy}}{\end{list}}

\newcounter{itemlistc}
\newcounter{romanlistc}
\newcounter{alphlistc}
\newcounter{arabiclistc}

\newcommand{\fcaption}[1]{
        \refstepcounter{figure}
        \setbox\@tempboxa = \hbox{\footnotesize Fig.~\thefigure. #1}
        \ifdim \wd\@tempboxa > 5in
           {\begin{center}
        \parbox{5in}{\footnotesize\smalllineskip Fig.~\thefigure. #1}
            \end{center}}
        \else
             {\begin{center}
             {\footnotesize Fig.~\thefigure. #1}
              \end{center}}
        \fi}

\newcommand{\tcaption}[1]{
        \refstepcounter{table}
        \setbox\@tempboxa = \hbox{\footnotesize Table~\thetable. #1}
        \ifdim \wd\@tempboxa > 5in
           {\begin{center}
        \parbox{5in}{\footnotesize\smalllineskip Table~\thetable. #1}
            \end{center}}
        \else
             {\begin{center}
             {\footnotesize Table~\thetable. #1}
              \end{center}}
        \fi}

\def\@citex[#1]#2{\if@filesw\immediate\write\@auxout
	{\string\citation{#2}}\fi
\def\@citea{}\@cite{\@for\@citeb:=#2\do
	{\@citea\def\@citea{,}\@ifundefined
	{b@\@citeb}{{\bf ?}\@warning
	{Citation `\@citeb' on page \thepage \space undefined}}
	{\csname b@\@citeb\endcsname}}}{#1}}

\newif\if@cghi
\def\cite{\@cghitrue\@ifnextchar [{\@tempswatrue
	\@citex}{\@tempswafalse\@citex[]}}
\def\citelow{\@cghifalse\@ifnextchar [{\@tempswatrue
	\@citex}{\@tempswafalse\@citex[]}}
\def\@cite#1#2{{$\null^{#1}$\if@tempswa\typeout
	{IJCGA warning: optional citation argument 
	ignored: `#2'} \fi}}

\def\pmb#1{\setbox0=\hbox{#1}
	\kern-.025em\copy0\kern-\wd0
	\kern.05em\copy0\kern-\wd0
	\kern-.025em\raise.0433em\box0}

\def\fnt#1#2{\footnotetext{\kern-.3em
	{$^{\mbox{\scriptsize #1}}$}{#2}}}

\def\thefootnote{\fnsymbol{footnote}}
\def\@makefnmark{\hbox to 0pt{$^{\@thefnmark}$\hss}}	
	
\def\ps@myheadings{%
    \let\@oddfoot\@empty\let\@evenfoot\@empty
    \def\@evenhead{\slshape\leftmark\hfil}
    \def\@oddhead{\hfil{\slshape\rightmark}}
    \let\@mkboth\@gobbletwo
    \let\sectionmark\@gobble
    \let\subsectionmark\@gobble
    }
\font\tenrm=cmr10
\font\tenit=cmti10 
\font\tenbf=cmbx10
\font\bfit=cmbxti10 at 10pt
\font\ninerm=cmr9
\font\nineit=cmti9

\font\eightrm=cmr8

\def\qed{\hbox{${\vcenter{\vbox{			
   \hrule height 0.4pt\hbox{\vrule width 0.4pt height 6pt
   \kern5pt\vrule width 0.4pt}\hrule height 0.4pt}}}$}}

\renewcommand{\thefootnote}{\fnsymbol{footnote}}  

\begin{document}
\setlength{\textheight}{7.7truein}  

\thispagestyle{empty}

\markboth{\protect{\footnotesize\it Instructions for Typesetting
Manuscripts}}{\protect{\footnotesize\it Instructions for
Typesetting Manuscripts}}

\normalsize\textlineskip

\setcounter{page}{1}

\copyrightheading{}		

\vspace*{0.88truein}

\centerline{\bf SCALAR NON-LUMINOUS MATTER IN GALAXIES }
\vspace*{0.37truein}
\centerline{\footnotesize BYUNG JOO LEE and TAE HOON LEE\footnote{
thlee@physics.soongsil.ac.kr}}
\baselineskip=12pt
\centerline{\footnotesize\it Department of Physics, Soongsil University,
Seoul 156-743, Korea }

\vspace*{0.225truein}
\publisher{(received date)}{(revised date)}

\vspace*{0.21truein}
\abstracts{ As a candidate for dark matter in galaxies, we study an $SU(3)$ 
triplet of complex scalar fields which
are non-minimally coupled to gravity. In the spherically symmetric 
static spacetime where the flat rotational velocity curves of stars 
in galaxies can be explained, we find simple solutions of scalar fields with 
$SU(3)$ global symmetry broken to $U(1)\times U(1)$,
in an exponential scalar potential, which will be useful in a quintessence 
model of the late-time acceleration of the Universe. }{}{}



\vspace*{1pt}\textlineskip	
\section{Introduction}	
\vspace*{-0.5pt}
\noindent
It is well known that the {\it flat} rotational velocity curves(FRVC) of stars in 
galaxies suggest the existence of non-luminous(or dark) 
matter in the galactic halo. 
If the energy density of dark matter in the galactic halo is proportional 
to $1/r^2$, then one can account for the 
FRVC of stars in 
galaxies.\cite{ha-nu} In the global monopole solution found by Barriola and Vilenkin, 
\cite{vi} scalar fields with global $O(3)$ broken symmetry are minimally coupled 
to gravity
 and the background spacetime has deficits of angle. 
Nucamendi {\it et} {\it al.}\cite{nu} suggested 
that the global monopole solution could explain the 
FRVC in galaxies because its energy density is proportional to $1/r^2$. 
But Harari and Loust\'{o}\cite{ha} showed 
that the monopole core mass is negative and that there are no bound orbits.  

Matos, Guzm\'{a}n, and Ure\~{n}a-L\'{o}pez\cite{phi} obtained solutions for 
metric coefficients of the spacetime in which
the FRVC of stars in galaxies could be explained. 
In the spacetime they found a solution of the scalar field in a singlet in a exponential 
potential. 
However its sign is negative, $V\propto -e^{-\varphi}$,\cite{-phi} and 
it is necessarily opposite to that of the exponential potentials that
had been considered in 
quintessence cosmologies.
We generalize them to the 
theory of complex scalar fields with $SU(3)$ global symmetry, 
coupled non-minimally to gravity.  
Complex scalar fields with $U(1)$ global symmetry were considered by 
Boyle, Caldwell, and Kamionkowski,\cite{sp}
as "spintessence" models for dark matter and dark energy. 
Difficulty of this spinning complex scalar field to be dark energy was discussed
by Kasuya.\cite{ka}    
A new quintessence model in which scalar fields possess a global $O(N)$ internal 
symmetry was studied by Li, Hao, and Liu.\cite{on} Both Boyle
 {\it et} {\it al.}
and Li {\it et} {\it al.} considered spinning(time-dependent) scalar fields. 
We study in this paper complex scalar fields which depend on $r$ only.

We consider $SU(3)$ scalar fields which 
are non-minimally coupled to gravity. 
In the spherically symmetric 
static spacetime in which the 
FRVC of stars 
in galaxies can be explained, we find simple solutions of scalar fields with 
$SU(3)$ global symmetry broken to $U(1)\times U(1)$,
in an exponential scalar potential $V= exp(-\Phi^{* a}\Phi^a) $,
choosing a special 
value of non-minimal coupling parameter $\xi $ as in Eq. (32).
The potential will be useful in a quintessence 
model for the late-time acceleration of the Universe.\cite{ac} 

\setcounter{footnote}{0}
\renewcommand{\thefootnote}{\alph{footnote}}

\section{ $SU(3)$ Scalar Fields Non-Minimally Coupled to Gravity. }
\noindent
The action for complex scalar fields, with an $SU(3)$ global symmetry, 
which are non-minimally coupled to gravity is given by
\begin{equation}
S=-\int d^4 x \sqrt{-g}[g^{\mu \nu}\partial_{\mu}\Phi^{* a} \partial_{\nu}\Phi^{ a}+\xi R \Phi^{* a}\Phi^{a}+V (\Phi^{* a}\Phi^{a}) ],
\end{equation} 
where $\Phi^{a}$ $(a=1,2,3)$ are complex scalar fields, $R$ is a scalar curvature,
 $\xi$ is a non-minimal coupling parameter, and $V(\Phi^{* a} \Phi^a )$ is a 
scalar potential.
Using the variational method, we obtain following equations for scalar fields $\Phi^{a}$:
\begin{equation}
\frac{1}{\sqrt{-g}} \partial_{\mu}(\sqrt{-g} g^{\mu \nu} \partial \Phi^a )-\xi R \Phi^a -\frac{\partial V }{\partial \Phi^{* a}} =0.
\end{equation}
Using the standard definition of
the energy-momentum tensor
\begin{equation}
T_{\mu \nu}=-\frac{1}{\sqrt{-g}} \frac{\delta S}{\delta g^{\mu \nu}}, 
\end{equation}
we have
\begin{equation}
T_{\mu\nu}=
\partial_{\mu}\Phi^{* a} \partial_{\nu}\Phi^{ a}
-\frac{1}{2} g_{\mu \nu} [ g^{\alpha \beta} \partial_{\alpha}\Phi^{* a} \partial_{\beta}\Phi^{ a}+V (\Phi^{* a}\Phi^{a}) ]+\xi  \Phi^{* a}\Phi^{a} G_{\mu \nu},
\end{equation}
where $G_{\mu \nu}$ is the Einstein tensor 
\begin{equation}
G_{\mu \nu}=R_{\mu \nu}-\frac{1}{2}g_{\mu \nu} R 
\end{equation}
with Ricci tensor $R_{\mu \nu}$.

Assuming the spherically symmetric static spacetime with the line element
\begin{equation}
ds^2 =g_{\mu\nu} dX^\mu dX^\nu =-B(r) dt^2 + A(r) dr^2 +r^2 d\theta^2 +r^2 sin^2 \theta d\phi^2 ,
\end{equation}
and the hedgehog Ansatz for scalar fields,
\begin{equation}
\Phi^a =F(r) \frac{x^a}{r},
\end{equation}
with a complex function $F(r)$ and $r=(x^a x^a)^{1/2} =(x^2 +y^2 +z^2 )^{1/2}$,
we get the equation for the complex function, $F$:
\begin{equation}
F''+\frac{F'}{2}(\frac{B'}{B}-\frac{A'}{A}+\frac{4}{r})-\frac{2 AF}{r^2}-\xi A RF-A\frac{\partial V}{\partial F^* }=0,
\end{equation}
where $F'=\frac{d F}{dr}, F''=\frac{d^2 F}{dr^2}, ...$,
and the scalar curvature is given by
\begin{equation}
R=-\frac{B''}{AB}+\frac{B'}{2AB}(\frac{A'}{A}+\frac{B'}{B}) +\frac{r}{2A}(\frac{A'}{A}-\frac{B'}{B})  +\frac{2}{r^2} (1-\frac{1}{A}).
\end{equation}

 In next section we solve both the field equation in Eq. (8) and the  
Einstein's equation given by
\begin{equation}
G_{\mu\nu}=\kappa T_{\mu\nu},
\end{equation} 
or
\begin{equation}
R_{\mu\nu}=\kappa(T_{\mu\nu}-\frac{1}{2}g_{\mu\nu}g^{\alpha\beta}T_{\alpha\beta}),
\end{equation}
with $\kappa=8\pi G$.
Instituting Eq. (4) into the Einstein's equation given in Eq. (11), we have
\begin{equation}
R_{\mu\nu}=
\kappa (\partial_{\mu}\Phi^{* a}\partial_{\nu}\Phi^{ a}+\frac{1}{2}g_{\mu\nu} V+ 
\xi R_{\mu\nu} \Phi^{* a}\Phi^{a} ),
\end{equation}
which can be reduced to

\begin{equation}
R_{\mu\nu}(1-\kappa  \xi  \Phi^{* a}\Phi^{a})=\kappa (\partial_{\mu}\Phi^{* a}\partial_{\nu}\Phi^{ a}+\frac{1}{2}g_{\mu\nu} V ).
\end{equation}

When elements of Ricci tensor calculated from metric coefficients
 in Eq. (6) and the Ansatz for $\Phi^a $ in Eq. (7) are substituted into Eq. (13), 
we obtain the following equations:
\begin{equation}
(1-\kappa \xi F^* F)[\frac{B''}{2A}-\frac{B'}{4A}(\frac{A'}{A}+\frac{B'}{B})+\frac{B'}{rA}  ]=-\kappa\frac{B}{2} V, 
\end{equation}

\begin{equation}
(1-\kappa \xi F^* F)[-\frac{B''}{2B}+\frac{B'}{4B}(\frac{A'}{A}+\frac{B'}{B})+\frac{A'}{rA} ]=\kappa[{F'}^{* } F'+ \frac{A}{2} V], 
\end{equation}

\begin{equation}
(1-\kappa \xi F^* F)[1-\frac{1}{A}+\frac{r}{2A}(\frac{A'}{A}-\frac{B'}{B})]=\kappa[F^* F+\frac{r^2 }{2} V]. 
\end{equation}
The first in above equations is the ($t, t$)-component of Eq. (10), the second 
is ($r, r$)-component, and the last is ($\theta, \theta$)-(or equivalently, 
($\phi, \phi$)-)component, respectively.

\section{FRVC of Stars in Galaxies and Solutions for them  }
\noindent

Circular motions of stars in a galaxy are determined by the 
geodesic equations in 
the spacetime with metric coefficients in Eq. (6).
When we consider the case $\theta=\frac{\pi}{2}$, 
dividing Eq. (6) by $d{\tau}^2$ 
and defining 
\begin{equation}
 \dot{t}\equiv \frac{dt}{d\tau}, \, \dot{r}\equiv \frac{dr}{d\tau},\,
\dot{\phi}\equiv \frac{d\phi}{d\tau},
\end{equation}
we get the equation
\begin{equation}
G(r, \dot{r}, \dot{\phi}, \dot{t})\equiv -B(r){\dot{t}}^2 +A(r){\dot{r}}^2 
+r^2 {\dot{\phi}}^2 =-1,
\end{equation}
where it is used that $ds^2 =-d\tau^2 $ in the unit system with $c=1$.
Since $\frac{\partial G}{\partial \phi}=0$ and
$\frac{\partial G}{\partial t}=0$, we have constants of motion:
\begin{eqnarray}
\frac{\partial G}{\partial \dot{\phi}} & \equiv &2r^2 \dot{\phi}=2L,  \\
\frac{\partial G}{\partial \dot{t}} & \equiv & -2B\dot{t}=-2E,
\end{eqnarray} 
where $E$ and $L$ are constants. 
Using Eqs. (19) and (20), Eq. (18) reads
\begin{equation}
{\dot{r}}^2 +V_{rad.}(r)=0,
\end{equation}
where
\begin{equation}
V_{rad.}(r)=\frac{1}{A(r)}(1+\frac{L^2}{r^2}-\frac{E^2}{B(r)}).
\end{equation}
Conditions for stars in a galaxy to have circular motions are\cite{nu}

\begin{eqnarray}
\dot{r}&=&0, \nonumber \\
\frac{\partial V_{rad.}}{\partial r}&=&0, \\
\frac{\partial^2 V_{rad.}}{\partial r^2}&>&0. \nonumber
\end{eqnarray}
From above equations, we get the rotational velocity 

\begin{equation}
v^{(\phi)} \equiv \frac{r}{\sqrt{B}}
\frac{ d \phi}{ dt}=\sqrt{\frac{rB'}{2B}}.
\end{equation}

The condition which the metric coefficient $B$ in Eq. (6) satify 
for the flat 
rotation curves of stars in galaxies to be explained is 
given by Guzm\'{a}n\cite{gu} and so on:\cite{xx}

\begin{equation}
\sqrt{\frac{rB'}{2B}}=v=constant.
\end{equation}
A solution to the above equation is
\begin{equation}
B(r)=B_o (\frac{r}{r_o})^{2 v^2}.
\end{equation}
We note that the metric coefficient function $B(r)$  
in Eq. (26) is determined independently of the other function $A(r)$ in Eq. (6), 
and that $B$ becomes singular for $r \rightarrow \infty$ 
and for $r \rightarrow  0$, but it is not seriously problematic since 
$v$ is very small(of the order $10^{-4} -10^{-3}$)\cite{ha-nu} and
Eq. (26) is valid only 
for the region where the galactic halo exists ($r_i<r<r_o$). For example, for our galaxy,
$r_o \simeq 230 $ Kpc,\cite{23} and for others, $r_o > 400$ Kpc.\cite{40}
Centers of galaxies are much more complicated 
and therefore $B$ in Eq. (26) should be substituted by other functions
for the region $r<r_i$.
 
When Eq. (26) is used, Eq. (14) reads

\begin{equation}
\frac{A'}{A}=\frac{2(v^2 +1)}{r}+\kappa\frac{rAV}{v^2(1-\kappa \xi F^* F)},
\end{equation}
which can be solved with the actual form of a potential $V(F^* F)$.

In a very simple case when 
\begin{equation}
V\simeq 0,
\end{equation}
Eq. (27) can be easily integrated and we get
\begin{equation}
A(r)\simeq  A_o(\frac{r}{r_o})^{2v^2 +2}.
\end{equation}
With $B(r) $ and $A(r)$ given in Eq. (26) and Eq. (27), 
respectively, the following funcion $F$ of scalar 
fields satisfies scalar field equations and Einstein's equation:
\begin{equation}
F=\eta\, exp({i \sqrt{4v^2 +2}} \,\, ln\, r   ),
\end{equation}
when 
\begin{equation}
\eta = constant
\end{equation}
 and 
\begin{equation}
\xi\simeq  -1.
\end{equation}

Actually the above condition in Eq. (28) can be satisfied if we consider the potential
\begin{equation}
V=exp(-\Phi^{* a}\Phi^a)=exp({-F^* F})=e^{-\eta^2},
\end{equation}
for $\eta \rightarrow \infty$,
which is natural in quintessence models for the late-time acceleration
of the Universe.\cite{ac} 

\section{Summary and Discussions }
\noindent

In the spherically symmetric static spacetime, 
we have determined metric coefficients 
to explain the 
FRVC of stars in galaxies, with a simple
scalar potential $V\simeq 0$. With the metric coefficients in Eq. (26)
and Eq. (29)
we have solved 
equations of scalar fields non-minimally coupled to gravity
and got simple solutions given in Eq. (7) and Eq. (30),
which have $SU(3)$ global symmetry broken 
to $U(1)\times U(1)$. 
If this $SU(3)$ is the flavor symmetry of light quarks, $u$, $d$, and $s$, then
the $SU(3)$ triplet of complex scalar fields we have considered as 
a candidate for galactic
dark matter might be mesons, $Q \bar{q}_a $, which are bound states of a heavy quark $Q$ and light 
anti-quark $\bar{q}_a$, where $Q$ is one of heavy quarks, $t$, $b$, 
or $c$ quark, and $\bar{q}_1 =\bar{u}$, $\bar{q}_2 =\bar{d}$ and $\bar{q}_3 =\bar{s}$.
 
Since
the metric function $B(r)$ given in Eq. (26) is valid only for $r_i <r<r_o$,
it will be interesting to build more realistic models which are valid 
for all distance from the origin, 
$0 \leq r<\infty$, and astronomical data of rotational velocities  
of stars
in galaxies can be compared with these: Beyond the region to which the galactic halo is pervaded,
$r>r_o$, $B(r)$ should be substituted by, for example, one of Schwarzschild spacetime,
and for the region $0\leq r<r_i$, $B(r)$ by a more appropriate function for explaining
 complicated dynamics of centers of galaxies. These functions will
have to be matched with $B(r)$ in Eq. (26) at boundaries, $r=r_i $ and $r=r_o$.

The assumption of Eq. (28) for a scalar potential can be realized when we consider 
an exponential potential as in Eq. (33). Exponential scalar potentials
would be very useful in cosmological 
models for the recently observed cosmic acceleration.\cite{ac} 
The late-time acceleration of the Universe could be explained 
by scaling solutions for the scale factor of the Universe,
driven by exponential potentials of quintessence.\cite{sc}
It thus will be more interesting to construct an cosmological model of the
accelerating Universe, by using our solutions of scalar fields with the exponential
potential in Eq. (33).

\nonumsection{Acknowledgements}
\noindent

T. H. Lee thanks Drs. P. Oh, J. Lee, J. Yoon, and J. M. Kim for useful discussions.  
This work was supported by Korea Research Foundation Grant(KRF-2001-015-DP0091).

\nonumsection{References}

\eject


\noindent

\end{document}